\documentclass[twocolumn,showpacs,preprintnumbers,prb]{revtex4-1}
\usepackage{amssymb}
\usepackage[dvips]{color}
\usepackage{epsfig}
\usepackage{dcolumn}
\usepackage{bm}
\usepackage{graphicx,color}

\def\g{\gamma}
\def\G{\Gamma}
\def\d{\delta}

\def\e{\epsilon}

\def\w{\omega}

\def\s{\sigma}

\begin{document}

\title{Thermal Effects on Photon-Induced Quantum Transport in a Single Quantum Dot}

\author{M. O. Assun\c{c}\~ao, E. J. R. de Oliveira, J. M. Villas-B\^oas, and F. M. Souza}
\affiliation{Instituto de F\'isica, Universidade Federal de Uberl\^andia, 38400-902, Uberl\^{a}ndia, MG, Brazil} 

\begin{abstract}
We theoretically investigate laser induced quantum transport in a 
single quantum dot attached to electric contacts. Our approach, based on
nonequilibrium Green function technique, allows to include thermal effects
on the photon-induced quantum transport and excitonic dynamics, enabling the study of non-Markovian effects. By solving a set of
coupled integrodifferential equations, involving correlation and propagator functions, we obtain the photocurrent and the dot occupation as a function
of time. Two distinct sources of decoherence, namely, incoherent tunneling 
and thermal fluctuations, are observed in the Rabi oscillations. As temperature increases a
thermally activated Pauli blockade results in a suppression of these oscillations.
Additionally, the interplay between photon and thermal induced electron populations results in a switch of the current sign as time 
evolves and its stationary value can be maximized by tunning the laser intensity.
\end{abstract}

\volumeyear{year} \volumenumber{number} \issuenumber{number}
\eid{identifier}
\date[Date: ]{\today}
\maketitle

\section{Introduction}

Quantum transport in semiconductor quantum dots and molecular systems is a subject of intense study
nowadays.\cite{review1} These nanoscaled devices provide a formidable environment to study fundamental aspects of quantum physics,
involving many-body correlations and light-matter interaction in regimes out of equilibrium.\cite{hh08}
These systems have a great potential to form a new generation of 
optoelectronic devices based on the unique electronic structure that arises from the quantum confinement.
For instance, quantum dots can produce a wealth of visible colors depending upon its size, even
white light with relatively high efficiency\cite{ter12} and potential to integrability with nanoelectronics.\cite{hnsk12} 
Additionally, with the great technological advances in the manufacturing of semiconductor quantum dot
system, it became possible to coherent monitor and control electron populations in two-level systems via different 
pump-probe techniques.\cite{ths01,hk01,pb02,smv10} In all these experiments the main signature of 
quantum coherent nonlinearity is Rabi oscillations, which has no classical counterpart.
More recently, Rabi oscillations was also reported in organic light-emitting diode.\cite{drmc08}
Such coherent optical manipulations constitute a fundamental ingredient to quantum information processing
in solid state devices that use electron-spin or excitonic states as qubits.\cite{mef02} Interestingly, holes in semiconductor quantum
dots have been revealed as an alternative to electrons in the manufacturing of spin qubits.\cite{ag11}

It was originally demonstrated by Zrenner \emph{et al.}\cite{az02} that coherent Rabi oscillations in a
two-level quantum dot photodiode can be monitored by photocurrents. Additionally, it was proposed
that a photocurrent in a self-assembled quantum dot photodiode can become spin-polarized due to an effective
exchange interaction via biexciton state.\cite{jmvb07} This result points out the potentiality of the present
system to future spintronic devices. It was also observed that the double dot structures present the ability
to increase the coherence time of indirect excitons.\cite{hsb10}
Recently, thermal effects on the excitonic Rabi rotations in a quantum dot system were investigated experimentally.\cite{ajr10_1,ajr10_2}
It was evidenced acoustic phonons as the main source of damping of the Rabi oscillations. 

In the present work we analyze how the temperature of nearby contacts tunnel coupled to a single quantum dot
affects the coherent optical dynamics. Applying nonequilibrium Green function technique 
to a microscopic Hamiltonian model, we write a set of coupled integrodifferential equations that describes the coherent evolution of
the electron-hole populations in the dot, enabling the study of non-Markovian effects.
A resonant laser field drives the electron-hole dynamics and generates a photocurrent. 
We find two contributions to the current. The first one comes from electrons in the dot that tunnel 
to a contact. This current is positive and is mainly induced by the laser field. The second current component
is related to electrons in the reservoir that acquires enough thermal energy to tunnel into the
dot. This second contribution charges the dot, thus
generating additional features not yet reported in the literature. Our main findings include
the suppression of the Rabi oscillation as the temperature of the nearby contact increases, 
a negative photocurrent due to a backwards charge flow, and a maximum photocurrent value
achieved when the laser intensity is comparable to the mismatch between the excited
dot level and the contact chemical potential. Moreover, we analyze how the initial electron population
of the conduction band, which is controlled by the temperature, affects the Rabi oscillations.

\section{Model and Formulation}

Figure (1) illustrates the system considered. It is composed of a quantum dot 
attached to a left and to a right electron reservoirs in the presence of a source-drain bias
voltage. A laser field shines the dot, thus generating electron-hole
pairs in it. The electrons in the conduction band and the holes in the valence
band can tunnel out from the quantum dot to the left and to the right reservoirs, respectively. 
This results in a photocurrent signal in the system. In the experimental 
point of view this system can be implemented in a structure of the kind n-GaAs--i--Schottky contact, as described
in Ref.[\onlinecite{az02}]. Alternatively, a $p$-$i$-$n$ junction can also be applied as describe in Ref.[\onlinecite{pwf00}], with self-assembled
quantum dots in the intrinsic region.

\begin{figure}[ht]
\centering        
\includegraphics[width=1\linewidth]{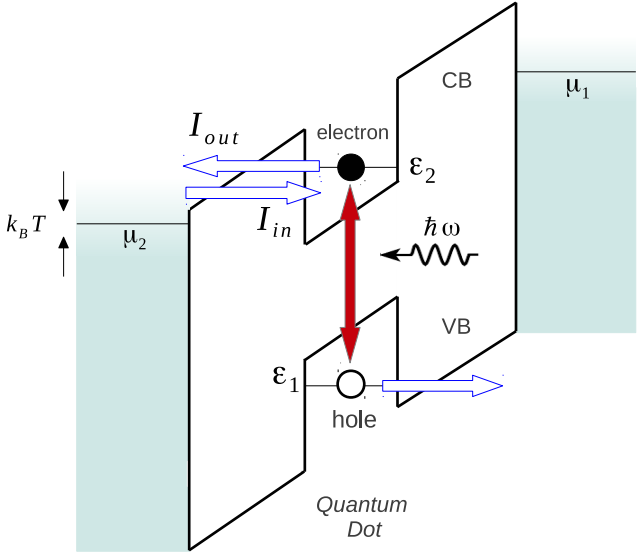}
\caption{(Color online) Sketch of the system studied. A quantum dot is tunnel coupled to 
both left and right reservoirs. The left reservoir illustrated has chemical potential
$\mu_2$ and temperature $T$. Due to the proximity of the
conduction band level $\e_2$ to $\mu_2$, the dot population thermally fluctuates. This induces nonlinearities
in the photocurrent driven by a laser field with resonant energy $\hbar \w$.} \label{fig1}
\end{figure}

The modeling Hamiltonian is given (per spin) by $H_\s=H_{D,\s}+H_{T,\s}+H_{L,\s}+H_{R,\s}$, with
\begin{equation}
 H_{D,\s}=\sum_{i}\e_{i\s} d_{i\s}^\dagger d_{i\s}+\g e^{-i \w t} d_{2\s}^\dagger d_{1\s} + \g^\star e^{i \w t} d_{1\s}^\dagger d_{2\s},
\end{equation}
where $\e_{i\s}$ is the dot level for spin $\s$ in the valence ($i=1$) or the conduction ($i=2$) band. The operators
$d_{i\s}$ ($d_{i\s}^\dagger$) annihilates (creates) one electron in level $i$ with spin $\s$. The parameter
$\g$ gives the optical transition between valence band and conduction band in the quantum dot.
This parameter can be controlled by the intensity of the incident radiation.
In our model electron-electron interaction is not accounted for in order to allow
analytical treatment.
To couple the dot to fermionic reservoirs we use the tunneling Hamiltonian
\begin{equation}
H_{T,\s}= \sum_i \sum_{k_i} (t_i c_{k_i\s}^\dagger d_{i\s} + t_i^* d_{i\s}^\dagger c_{k_i \s}), 
\end{equation}
where $c_{k_i \s}$ ($c_{k_i \s}^\dagger$) annihilates (creates) one electron in the right ($i=1$)
or the left ($i=2$) lead.\cite{comment4} 
The parameter $t_i$ gives the dot-leads coupling strength. 
Finally, the free-electron energies of the electrons in both leads are given by
\begin{equation}
 H_{L,\s}+H_{R,\s}=\sum_{i=1}^2  \sum_{k_i} \e_{k_i \s} c_{k_i\s}^\dagger c_{k_i\s}.
 \end{equation}
In the present model we assume that the tunneling rates are larger than spontaneous
emission rates, so that electron-hole recombination will be neglected.

Our main task is to explore the effects of the reservoirs temperature on the Rabi oscillations
and the photocurrent. To this goal we must find the lesser and retarded Green functions
of the quantum dot, i.e., $G_{ij\s}^<(t,t)=i \langle d_{j\s}^\dagger(t) d_{i\s}(t) \rangle$ and $G_{ij\s}^r(t,t')=-i \theta(t-t') \langle \{ d_{i\s}(t),d_{j\s}(t') \} \rangle$.
Note that the occupations of the levels $\e_{i\s}$ are given by $n_{i\s}(t)=\mathrm{Im} \{G_{ii\s}^<(t,t)\}$, while the photocurrent is defined
as $I_{i\s}=-e \langle \dot{N}_{i\s} \rangle = -e i \langle [H_\s,N_{i\s}] \rangle$, ($\hbar=1$) with $N_{i\s}=\sum_{k_i} c_{k_i \s}^\dagger c_{k_i \s}$ being the total 
number of particles operator. Following Ref. [\onlinecite{hh08}] one can show that
\begin{equation}\label{I2s}
 I_{2\s}(t)=-e \G_{2\s} n_{i\s}(t) + 2 e \mathrm{Re} \{\Phi_{22\s}^r \}
\end{equation}
where $\Phi_{22\s}^r(t)=\int_{-\infty}^t dt_1 G_{22\s}^r(t,t_1) \phi_{2\s}(t_1,t)$ and
\begin{equation}
 \phi_{i\s}(t_1,t)= i \G_{i\s} \int \frac{d\e}{2\pi} f_i(\e) e^{-i\e(t_1-t)},
\end{equation}
with $i=1,2$. Here $f_i(\e)$ is the Fermi function to $i$-th reservoir and $\G_{i\s}=2\pi |t_i|^2 \rho_{i\s}$ is the
tunneling rate with $\rho_{i\s}$ being the density of states of the corresponding reservoir for spin component $\s$.
The present formalism allows the inclusion of ferromagnetic leads by considering spin-dependent tunneling rates
$\G_{i\s}$.\cite{fms04} According to Eq. (\ref{I2s}) the current at time $t$ has two contributions, one that is instantaneous
and proportional to the dot occupation $n_{i\s}(t)$ ($I_{out}$) and a second one that involves the whole history of the system ($I_{in}$).
In this second term, a time integral of the correlation function $G_{22\s}^r(t,t_1)$ weighted by a thermal dependent
function $\phi_{2\s}(t_1,t)$ should be carried on, ranging from $-\infty$ to the present time. 
All the thermal effects arise via this memory integral.
This contrasts to the density matrix approach used in quantum optics formulation that 
in general does not account for thermal and 
memory effects in the standard Markov approximation.\cite{jmvb05,dm08}

Calculating the time derivative of $G_{ij\s}^<(t,t)$ we arrive at
\begin{equation}\label{dGlesserdt}
 i\frac{\partial}{\partial t}\mathbf{G}_\s^<(t,t)=\mathbf{M}_\s (t) \mathbf{G}_\s^<(t,t) - \mathbf{\Phi}_\s(t),
\end{equation}
where the lesser Green function is written in a vector-like form $\mathbf{G}_\s^<=[G_{11\s}^<,G_{12\s}^<,G_{21\s}^<,G_{22\s}^<]^T$ and 
$\mathbf{\Phi}_\s=[\Phi_{11\s}^\prime,\Phi_{12\s}^\prime,\Phi_{21\s}^\prime,\Phi_{22\s}^\prime]^T$. Here 
$\Phi_{ij\s}^\prime(t)=\Phi_{ij}^r(t)-\Phi_{ij}^a(t)$, with 
\begin{equation}
 \Phi_{ij}^{r}(t)=\int_{-\infty}^t dt_1 G_{ij\s}^{r}(t,t_1) \phi_{j\s}(t_1,t),
\end{equation}
and
\begin{equation}
 \Phi_{ij}^{a}(t)=\int_{-\infty}^t dt_1 \phi_{i\s}(t,t_1) G_{ij\s}^{a}(t_1,t).
\end{equation}

The matrix in Eq. (\ref{dGlesserdt}) is given by


\begin{equation}
 \mathbf{M}_\s(t)=\left[
\begin{array}{cccc}
  -i \G_{1\s} & -\g e^{-i \w t} & \g e^{i \w t} & 0 \\
  -\g e^{i \w t} & \w_{12}-\frac{i}{2}\G_{\s} & 0 & \g e^{i \w t} \\
  \g e^{-i \w t} & 0 & \w_{21} -\frac{i}{2}\G_{\s} & -\g e^{-i \w t} \\
  0 & \g e^{-i \w t} & -\g e^{i \w t} & -i \G_{2\s}
\end{array}
\right],
\end{equation}
with $\w_{ij}=\e_i-\e_j$ and $\G_\s=\G_{1\s}+\G_{2\s}$. It is yet valid to note that in the absence of the reservoirs, 
Eq. (\ref{dGlesserdt}) reduces to the well known semiconductor Bloch equations.\cite{hh08} In order to
determine $\mathbf{\Phi}_\s(t)$ we need the retarded and advanced Green functions $G^{r,a}_{ij}(t,t')$.
Taking the time derivative with respect to $t'$ we obtain
\begin{equation}\label{dGrdt}
 - i \frac{\partial}{\partial t'}\mathbf{G}_\s^r(t,t')=\d(t-t') \left[
\begin{array}{c}
   \chi_{+} \\
   \chi_{-}
\end{array}
\right]
 +\mathbf{P}_\s(t') \mathbf{G}_\s^r(t,t'),
\end{equation}
where $\mathbf{G}_\s^r=[G_{11\s}^r,G_{12\s}^r,G_{21\s}^r,G_{22\s}^r]^T$, $\chi_+$ and $\chi_-$ are the two-component
Pauli spinors, $\chi_+=[1,0]^T$ and $\chi_-=[0,1]^T$, and the matrix $\mathbf{P}_\s(t')$ is 
defined according to
\begin{equation}
 \mathbf{P}_\s(t)=\left[
\begin{array}{cccc}
  \d_1 & \g e^{-i \w t'} & 0 & 0 \\
  \g e^{i \w t'} & \d_2 & 0 & 0 \\
  0 & 0 & \d_1 & \g e^{-i \w t'} \\
  0 & 0 & \g e^{i \w t'} & \d_2
\end{array}
\right],
\end{equation}
with $\d_l=\e_l-\frac{i}{2}\Gamma_{l\s}$. Solving numerically Eqs. (\ref{dGlesserdt}) and (\ref{dGrdt}) we 
obtain the occupation $n_{i\s}(t)$ and the photocurrent. In what follows we
present our results.
               
\section{Parameters} 

In order to keep the generality of our results, we express the time in units of $t_0=\hbar/\Gamma_0$,
where $\Gamma_0$ is proportional to the tunneling rate between dot and reservoirs. For simplicity we
assume the wideband limit, where the tunneling rates are energy independent and we 
set $\Gamma_{1\s}=\Gamma_{2\s}=\Gamma_0$. The current unit is given by $I_0=e\Gamma_0/\hbar$ and all energies are in units of 
$\Gamma_0$.\cite{comment8} Experimentally, $\G_0$ depends on the tunnel barrier and it can be easily controlled by an external gate voltage.
We find for quantum dot systems $\Gamma_0 \sim 1 \mu eV - 100 \mu eV$,\cite{dgg98_1,dgg98_2,fs99} which results in $I_0 \sim 0.24 nA - 24 nA$.\cite{comment7} 
The time $t_0$ ranges in the interval $t_0 \sim 6.5 ps$ ($\Gamma_0 = 100 \mu eV$) - $0.65 ns$ ($\Gamma_0 = 1 \mu eV$).\cite{comment6}
Additionally, the quantum dot valence and conduction band levels are given by $\e_{1\s}=\e_1=-100\Gamma_0$ and $\e_{2\s}=\e_2=2\Gamma_0$, respectively.\cite{comment3} 
Both levels are measured with respect to the chemical potential $\mu_2=0$, which is taken as our energy reference.
The chemical potential of the right side $\mu_1$ is given by $\mu_1-\mu_2=eV_{b}$, where $V_{b}$ is the bias voltage.
In what follows we adopt $eV_{b}=10\Gamma_0$.
Finally, we assume $k_B T \sim 0.1 \Gamma_0 - 2 \Gamma_0$ and $\gamma = 7 \Gamma_0$,\cite{comment5} except when those parameters explicitly change in the plots.

\section{Results}

\begin{figure}[ht]
\centering        
\includegraphics[width=0.9\linewidth]{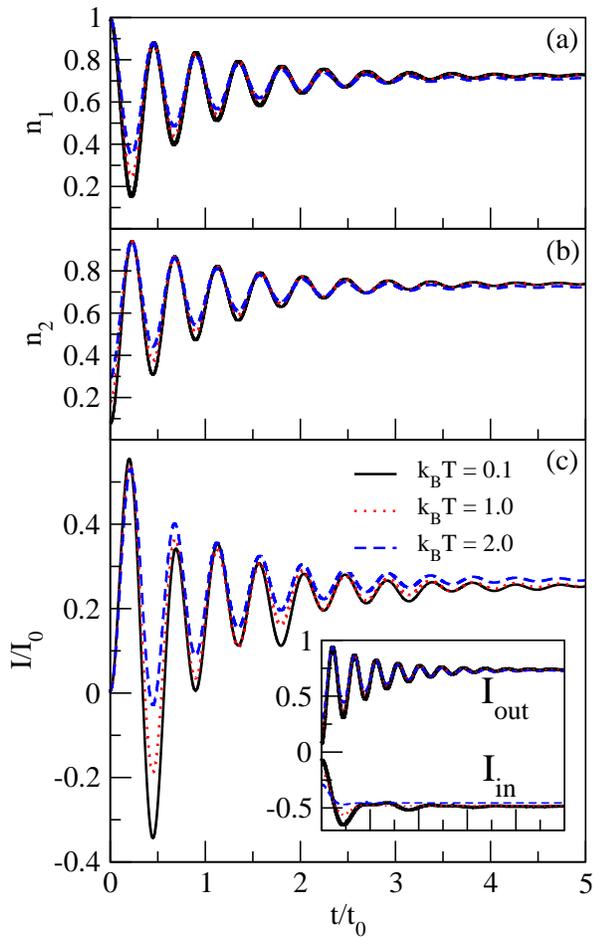}
\caption{(Color online)(a)-(b) Occupation of the levels $\e_1$ and $\e_2$ and (c) photocurrent as a function
of time for differing temperatures. As time evolves all these quantities exhibit coherent 
Rabi oscillations. The amplitude decreases with time due to the incoherent tunneling
between dot and electrodes. The oscillation amplitude is also suppressed as $k_B T$ increases
due to a thermal induced Pauli blockade. For $t \approx 0.5 t_0$ the photocurrent is 
dominated by a backwards current, thus becoming negative. In the inset we show separately the in 
and out current components.} \label{fig2}
\end{figure}

Figure \ref{fig2}(a)-(b) shows the evolution of the electron and hole occupations in the quantum dot
for differing temperatures $k_B T$. At $t=0$ the valence band level is fully occupied with $n_1=1$ while 
the occupation of the conduction band is $n_2 \approx 0.1$.\cite{foot1} This small occupation comes from the proximity of
the level $\e_2$ to the Fermi energy of the left reservoir, which allows thermal excited electrons to tunnel
into the dot. Initially ($t=0$) the quantum dot occupation
is calculated using the equation
\begin{equation}\label{ni}
 n_{i\s}(t=0)=\int \frac{d\e}{2\pi i} G_{ii\s}^<(\e),
\end{equation}
where the lesser Green function is given by the Keldysh equation $G_{ii\s}^<(\e)=G_{ii\s}^r(\e) \Sigma^<_{i\s}(\e) G_{ii\s}^a(\e)$,
where $G^{r(a)}_{ii\s}$ is the retarded (advanced) Green function of the dot attached to the leads without laser field
and $\Sigma_{i\s}^< = i \Gamma_{i\s} f_i$. 
As $k_B T$ increases, electrons in the
left electrode acquire enough thermal energy to enhance the population $n_2$ at $t=0$, while
$n_1$ remains the same due to $\e_1 \ll \e_F$. When the system starts to evolve
in the presence of a laser field, the occupations $n_1$ and $n_2$ develop the
characteristic Rabi oscillations. In the small temperature regime these oscillations are more pronounced
for small times and become suppressed as the time increases. 
This is due to the decoherence imposed by the tunneling between dot and reservoirs. 
For large enough times both $n_1$ and $n_2$ reach constant values.

As the temperature increases, the amplitude of the Rabi oscillations shrinks for all times. 
This is directly related to the enhancement of the initial population $n_2$ with temperature. With the level $\e_2$
becoming more populated, the Pauli exclusion principle makes it more difficult to one electron
with the same spin in the valence band ($\e_1$) to jump to the conduction band ($\e_2$).
So we observe two sources of suppression to the coherent Rabi oscillations: 
\emph{(i)} coupling to reservoirs and \emph{(ii)}  thermal activated Pauli blockade. 

The photocurrent seen in Fig. \ref{fig2}(c), at least to some extent, reflects the $n_2$ behavior. It oscillates in time with a
decreasing amplitude, tending to a stationary nonzero value. 
Additionally, the amplitude of the Rabi oscillations is also reduced as $k_B T$ increases,
following the behavior of $n_2$.
Interestingly, for small enough temperatures and shorter times,
the photocurrent oscillations attain negative values, which corresponds to
an unusual flow of electrons from the reservoir into the dot (see solid line, $k_B T = 0.1\G_0$, around $t=0.5 t_0$).
In order to gain further insight of this effect, in the inset of Fig. \ref{fig2}(c) we show separately the current components $I_{out}$ and $I_{in}$.
The outgoing current is positive which means that electrons are flowing from the dot to the reservoir.
The incoming current gives a negative contribution to the current, which corresponds to electrons flowing in the opposite way, i.e.,
from the electrode into the dot. 
Around $t=0.5t_0$, the $I_{in}$ component presents a dip, which pulls down the total photocurrent, making it negative.
When $k_B T$ increases, this dip is suppressed and the photocurrent assumes positive values for all times.

\begin{figure}[ht]
\centering        
\includegraphics[width=0.9\linewidth]{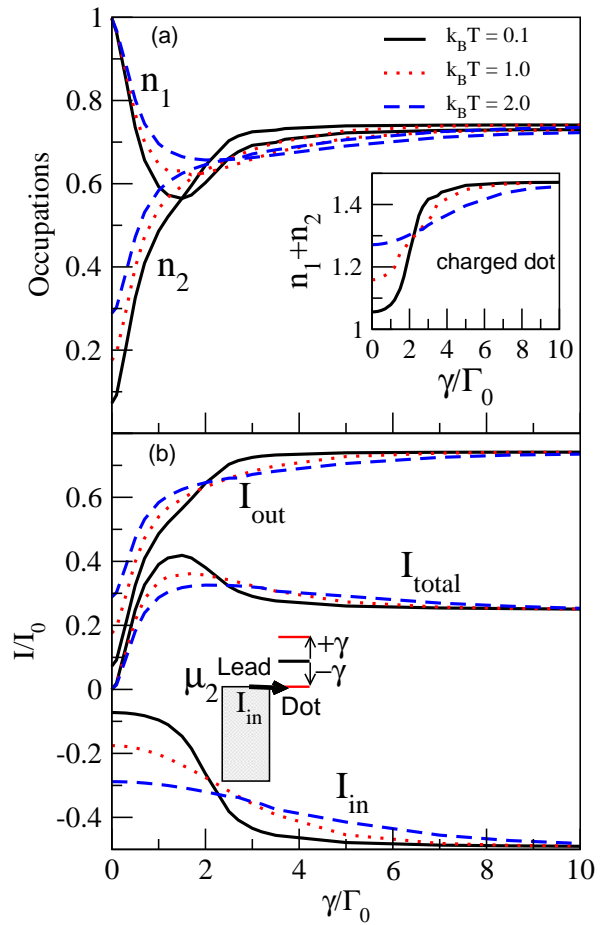}
\caption{(Color online)(a) Occupations $n_1$ and $n_2$ and (b) current components against $\gamma$. While
$n_2$ increases monotonically with $\g$, $n_1$ is initially suppressed, reaches a minimum and then increases
slightly. Oppositely, the photocurrent increases with $\g$, reaches a maximum and then becomes slightly
suppressed. This nonmonotonic behavior in both $n_1$ and $I$ can be understood looking at the current
component $I_{in}$. When $\g \approx \e_2-\mu =2 \G_0$, electrons in the reservoir can tunnel
into the dot, thus populating $\e_2$. This enhancement (in modulus) of the backwards current suppresses
the photocurrent for $\g > 2 \G_0$ and increases further $n_2$. In the inset we plot $n_1+n_2$. Note that
the dot charges for $\g > 2 \G_0$.} \label{fig3}
\end{figure}

Figure \ref{fig3}(a) shows how $n_1$ and $n_2$ evolves as a function of the parameter $\g$,
for differing temperatures. For all $k_B T$ values we observe $n_1 = 1$ for $\g=0$, while $n_2$ increases with
$k_B T$ for $\g=0$. This enhancement of $n_2$ with temperature comes from the thermal excited electrons in the reservoir
that acquires enough energy to jump into the dot as $k_B T$ increases. Both $n_1$ and $n_2$ are obtained
via Eq. (\ref{ni}) for $\gamma = 0$. In the presence of the laser field, $n_2$ increases monotonically with $\gamma$,
while $n_1$ is initially suppressed and then it is enhanced, thus developing a minimum around $\g \approx \e_2 -\mu_2=2 \G_0$.
Note also that $n_2$ presents a further enhancement near $\gamma = 2 \G_0$. 
In the inset of Fig. \ref{fig3}(a) we show the sum $n=n_1+n_2$ against $\gamma$. When the resonant
condition $\gamma = \e_2 - \mu_2$ is attained, the total occupation presents a steeper enhancement for low temperature.
For larger $k_B T$ a more broaden increasing is found. It is valid to note that the sum $n_1+n_2$ is
not limited to one, as expected in a standard two-level system with one level being initially occupied and the other one 
being initially empty. Here the level $\e_2$ is not populated only by the laser field, but also by the left reservoir.
The occupation profiles will be more clearly understood looking at the current components
in the next plot.

In Fig. \ref{fig3}(b) we plot separately the current components $I_{out}$, $I_{in}$ and the total current $I=I_{out}+I_{in}$.
While the outgoing current follows $n_2$, the incoming current is strongly increased (in modulus) around
$\g \approx \e_2 - \mu_2$. As a result, the photocurrent is suppressed due to this backwards current,
thus developing a peak close to $\gamma = \e_2 - \mu_2$. Increasing even further the temperature,
the thermal fluctuations of the reservoirs yield to a more effective injection of electrons 
into the dot. This makes $I_{in}$ starts at higher absolute values for $\gamma \lesssim 2\G_0$.
This amplification of $I_{in}$ suppresses the photocurrent when compared to its low temperature profile.
In the presence of a laser field in resonance with the difference $\e_2 - \e_1$, doublets
emerge in the spectrum of the system,\cite{sha55,brm69} as illustrated in the drawn of Fig. \ref{fig3}(b).\cite{comment2}
As the laser intensity increases, the lower energy peak of the doublet eventually 
attain resonance with the reservoir chemical potential at $\gamma = \e_2 -\mu_2$. This allows electrons to 
resonantly tunnel from the lead into the dot, thus generating a backwards current that suppresses the total photocurrent
and increases the $n_2$ population. When the lower peak lies below $\mu_2$, the enhancement of $k_B T$ tends to depopulate this channel,
consequently suppressing $I_{in}$, as seen in Fig. \ref{fig3}(b) for $\g \gtrsim 2 \G_0$.

\begin{figure}[ht]
\centering        
\vskip0.5cm
\includegraphics[width=0.9\linewidth]{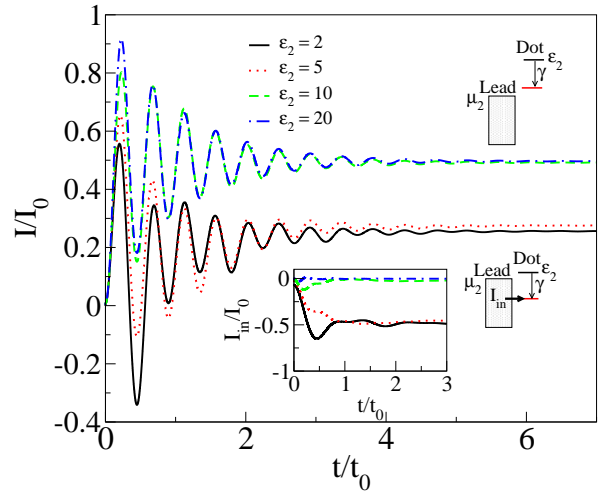}
\caption{(Color online) Photocurrent against time for different $\e_2$ values, with respect to $\mu_2$. As $\e_2$ increases the photocurrent
is amplified. This is due to the suppression of the incoming current for larger $\e_2$. The temperature adopted is $k_B T = 0.1 \G_0$.
In the inset we show the incoming current component for all the $\e_2$ values used. As $\e_2$ enlarges, the
incoming current tends to zero.} \label{fig4}
\end{figure}

Finally, it is valid to point out that the $I_{in}$ current component plays a role in the transport
whenever $\e_2-\gamma \leq \mu_2$, which allows electrons in the reservoir to resonant tunnel
to the dot. It is possible to entirely suppress the incoming current by
moving $\e_2$ high enough above $\mu_2$, so that $\e_2-\gamma > \mu_2$.
For this regime electrons can flow only in one direction, i.e., from the
dot to the reservoir, thus reducing the backwards charge flow.
Fig. \ref{fig4} shows the photocurrent against time for different $\e_2$ values. 
For $\e_2=2\G_0$ and $5\G_0$ the channel $\e_2-\g$ lies below $\mu_2$, as illustrated
in the energy diagram at the lower part of the panel. This results in a relatively high $|I_{in}|$ 
component (see the inset). On the contrary, for $\e_2=10\G_0$ and $20\G_0$ we find $\e_2-\g$ higher than
$\mu_2$ (see the upper energy diagram sketched), which suppresses $|I_{in}|$
and makes the total current larger.

\section{Conclusion}

In conclusion, via nonequilibrium Green function technique we have investigated the dynamics of electron-hole
pairs in a quantum dot tunnel coupled to Fermionic reservoirs. We found that the thermal fluctuation of the reservoir and 
consequent occupation of the conduction band level appears as a new source of decoherence for the optically induced Rabi oscillation in QDs, 
which has not been reported yet. As temperature increases, the thermal excited carriers 
in the left reservoir acquires enough energy to tunnel into the dot. This gives rise to an enhancement
of electronic dot population, which results in a thermal activated Pauli blockade that suppresses slightly the Rabi oscillations.
This effect is strongly dependent on the temperature of the reservoirs and on the mismatch 
between $\e_2$ and $\mu_2$. Finally, a nonlinearity signature is found in the current against $\g$. 
This results from a doublet that brings into resonance a transport
channel with the chemical potential $\mu_2$. This laser induced resonance generates a competition
between outgoing and incoming currents in the quantum dot that yields to the observed nonlinearities.
As a final remark, we note that the present study is a fundamental example of the use of 
nonequilibrium Green function for optical processes, which can be applied to more intricate systems, including non-Markovian processes.


The authors acknowledge the Brazilian agencies CNPq, CAPES and FAPEMIG for financial support.


\begin{thebibliography}{20}
\bibitem{comment8} The parameter $\Gamma_0$ in our model gives the relaxation rate due to incoherent tunneling processes.
The characteristic relaxation time is $\hbar/\Gamma_0$.
\bibitem{comment3} The parameter $\e_{1\s}=-100\G_0$ is in fact much smaller than the realistic values for semiconductor dots that result in excitonic
energy typically of $\e_2-\e_1 \approx 1$eV. However, since $\e_{1\s}$ is already much larger than any other
energy scale in the problem, if its value is increased even further no changes will be observed in the results.
\bibitem{comment5} It is valid to note that our formulation wouldn't be valid in the limit where $\gamma$ is comparable
to the difference $\e_2-\e_1$ or the laser frequency $\w$, since we adopt the rotating wave approximation. Additionally, in the experimental point of view
$\gamma$ is much smaller than $\e_2-\e_1$ or $\w$.
\bibitem{foot1} It is valid to point out that in the standard approach of the two-level semiconductor quantum dot systems, 
based on density matrix formalism, it is adopted the base $|0\rangle$ and $|1\rangle$ for no-exciton and one-exciton in the dot, respectively.
The initial state $|0\rangle$ in our formalism stands to $n_1 = 1.0$ and $n_2 = 0.0$, which means that the conduction band is empty, 
while the valence band is filled with one electron. 
Though, due to the thermal effects reported here, the occupation $n_2$ fluctuates, thus resulting in $n_2\neq0$.
\end{thebibliography}
\end{document}